\documentclass[preprint2]{aastex}

\usepackage{amsmath,amssymb,pgf,graphicx,color,url}
\shorttitle{Search for differences between radio-loud and radio-quiet gamma-ray pulsar populations with Fermi-LAT data}
\shortauthors{E.V. Sokolova, G.I. Rubtsov}
\begin{document}
\title{Search for differences between radio-loud and radio-quiet gamma-ray pulsar populations with Fermi-LAT data}
\slugcomment{INR-TH/2016-001}
\author{E.V. Sokolova and G.I. Rubtsov} 
\email{sokol@ms2.inr.ac.ru, grisha@ms2.inr.ac.ru}
\affil{Institute  for Nuclear Research of the Russian Academy of Sciences,
Moscow 117312, Russia}

\begin{abstract}
Observations by Fermi-LAT have enabled us to explore the population of
non-recycled gamma-ray pulsars with a set of 112 objects. It was
recently noted that there are apparent differences in the properties
of radio-quiet and radio-loud subsets. In particular, the average
observed radio-loud pulsar is younger than the radio-quiet one and is
located at smaller Galactic latitude. Even so, the analysis based on
the full list of pulsars may suffer from selection effects. Namely,
most radio-loud pulsars are first discovered in the radio band, while
radio-quiet ones are found using the gamma-ray data. In this work we
perform a blind search for gamma-ray pulsars using the Fermi-LAT data
alone, using all point sources from the 3FGL catalog as the
candidates. Unlike our previous work, the present catalog is
constructed with a semi-coherent method based on the time-differencing
technique and covers the full range of characteristic ages down to 1
kyr. The search resulted in a catalog of 40 non-recycled pulsars, 25
of which are radio-quiet. All pulsars found in the search were
previously known gamma-ray pulsars. We find no statistically
significant differences in age and Galactic latitude distributions for
the radio-loud and radio-quiet pulsars, while the rotation period
distributions are marginally different with statistical probability of
$4\times 10^{-3}$.  The fraction of radio-quiet pulsars is estimated
as $\epsilon_{RQ}=63\pm 8\%$. The results are in agreement with the
predictions of the outer magnetosphere models, while the Polar cap
models are disfavored.
\end{abstract}

\section{Introduction}

The number of known gamma-ray pulsars has grown rapidly since the
Fermi Large Area Telescope (LAT) started taking data in August
2008. At present
205\footnote{\url{https://confluence.slac.stanford.edu/display/GLAMCOG/Public+List+of+LAT-Detected+Gamma-Ray+Pulsars}}
gamma-ray pulsars are identified, including 112 non-recycled
pulsars. The latter includes 61 radio-loud pulsars and 51 radio-quiet
ones, see~\citet{Caraveo:2014lra,Grenier:2015pya} for a review. With the
only exception of Geminga~\citep{GemingaX,GemingaG}, the detection of
the radio-quiet pulsars became possible only with the sensitivity of
Fermi-LAT.  High-performance numerical methods were designed and
implemented based on the time-differencing technique
~\citep{Atwood2006,Abdo12,Parkinson69,Pletsch1,Pletsch2,Pletsch3}.

With the statistics in hands one may raise questions on the model of
the gamma-ray emission and corresponding mechanism of the
radio-quietness. Two general classes of the pulsar gamma-ray emission
models are discussed. The first class includes the so-called Polar cap
(PC) models~\citep{Sturrock:1971zc}. In these models gamma rays are
produced by electrons and positrons accelerated in the polar cap
region near the surface of the neutron star. In the PC models the
gamma-ray and radio beams are generally co-aligned. The latter is
considered narrower than the former and therefore some of the pulsars
are observed as radio-quiet~\citep{SturnerDermer:1996}. Moreover, in
the PC models the fraction of the radio-quiet pulsars depends on the
pulsar's age~\citep{Gonthier:2002}. In the second class of models the
gamma-ray emission is produced in the outer magnetosphere (OM) of the
pulsar~\citep{Cheng:1986qt,Perera:2013wza}. In the OM models the
radio-quietness finds a geometrical description as the gamma-ray and
radio-beams orientations are naturally diverse. Moreover, beam
geometry within the OM models may serve as a consistent description of
recent Chandra and XMM-Newton observations of gamma-ray
pulsars~\citep{Marelli}. The latter showed that on average radio-quiet
pulsars have higher gamma-ray-to-x-ray flux ratio than radio-loud ones.

It was noted that the observed fraction of the radio-quiet objects is
relatively small for young pulsars~\citep{Ravi:2010sm}. This
observation may be interpreted in terms of evolution of the radio-beam
solid angle~\citep{Ravi:2010sm,Watters:2010jb}. Alternatively this may
be an effect of the observational selection
bias~\citep{Caraveo:2014lra}. While radio-quiet pulsars are discovered
in a blind search with gamma-ray data, there are multiple ways to find
radio-loud pulsars. The latter may be found either in radio surveys or
with follow-up observations of gamma-ray sources. In most cases the
gamma-ray pulsations of radio-loud pulsars are found with ephemerides
from radio observations. Nevertheless there are pulsars with pulsed
radio-emission detected following the gamma-ray pulsations. 

In this {\it Paper} we construct
a less biased catalog of the gamma-ray selected pulsars by performing a
blind search for gamma-ray pulsars using the Fermi LAT data alone. The
search is more extensive with respect to our preceding blind
search~\citep{pscan1}. With the novel efficient semi-coherent method a
complete range of characteristic ages starting from 1 kyr is
covered. No radio or optical observation data are used.

The paper is organized as follows. Fermi-LAT data selection and
preparation procedures are explained in Section~\ref{sec:data}. The
implementation details of the semi-coherent method are given in
Section~\ref{sec:method}.  The catalog of the gamma-ray selected
pulsars, comparison of radio-quiet and radio-loud gamma-ray pulsar
populations and comparison of the results to the predictions of the
pulsar emission models are presented in Section \ref{sec:results}.

\section{Data}
\label{sec:data}
	
The paper is based on publicly-available weekly all-sky Fermi-LAT data
for the time period from 2008 August 4 till 2015 March 3 (Mission
elapsed time from 239557418 to 447055673) \citep{FermiLAT}\footnote{\url{http://fermi.gsfc.nasa.gov/ssc/data/access/}}. We select SOURCE class events from
Reprocessed Pass 7 data set with energies from 100~MeV to 300~GeV. The
standard quality cuts are applied using the {\it Fermi Science Tools
  v9r32p5} package. These include $100^\circ$ and $52^\circ$ upper
constraints for zenith angle and satellite rocking angle, respectively.

We search for pulsations using the location of each of 3008 point
sources from the Fermi LAT 4-year Point Source Catalog
(3FGL)~\citep{3FGL}. The requirement of blindness to everything except
gamma-ray emission binds us to the coordinates from the 3FGL
catalog. Although the precision of the gamma-ray source positioning is
one of the factors limiting sensitivity of the scan~\citep{Pletsch1}
we did not scan over the sky location due to computational
complexity. This has an effect on the sensitivity of the method to the
pulsars with the frequency higher than $16$\,Hz as shown
by~\citet{Dormody:2011}.

A model of the $8^\circ$ radius circle sky patch is constructed for
each of the candidates. The model includes galactic and isotropic
diffuse emission components and all 3FGL sources within $8^\circ$ from
the position of interest. We optimize the parameters of the model with
unbinned likelihood analysis by the {\it gtlike} tool. Normalization
of the galactic and isotropic diffuse emissions and spectral
parameters of the target source were considered as free parameters,
while the spectral parameters of all neighboring sources were fixed at
the 3FGL values. The spectral form for each source is power-law,
log-parabola or power-law with exponential cutoff based on the 3FGL
data. Next, using the {\it gtsrcprob} tool each photon is assigned a
weight - probability to originate from the given pulsar candidate.
For computational efficiency we keep 40000 events with the highest
weights for each source. Finally, the photon arrival times are
converted to barycentric frame using the {\it gtbary} tool.

\section{Method}
\label{sec:method}

The search for pulsations is performed with the time-differencing
technique first proposed by \citet{Atwood2006} and subsequently
refined by \citet{Pletsch1}. We scan over the pulsar's frequency $f$ and
spin-down rate $\dot{f}$ using the array of barycentric photon arrival
times $t_a$ and corresponding weights $w_a$.  First, the arrival times
are corrected to compensate for frequency evolution
\begin{equation}
\label{time}
\tilde{t}=t + \frac{\gamma}{2}(t-t_0)^{2}\,,
\end{equation}
where $\gamma = \dot{f}/f$ and $t_0=286416002$ (MJD 55225) is a
reference epoch. The spectral density $P_m$ is obtained with a Fourier
transform of the time differences

\begin{equation}
P_m=\mathrm{Re} \sum_{a,b=1}\Pi(\Delta t_{ab}/2T)w_aw_{b}e^{-2\pi
  if_{m}\Delta{t_{ab}}},
\end{equation}

where $\Delta{t_{ab}}=t_b-t_a$, $f_{m} = m/T$, $T=2^{19}$ and $\Pi(x)$
is the rectangular function which takes the value of 1 for $-0.5
<x<0.5$ and 0 otherwise. Technically we consider only positive time
differences and bin the time interval $(0,T)$ into $N = 2^{26}$
bins. Within the cycle over all event pairs one calculates the sum of
product of weights $w_aw_b$ in each bin. This prepares an input for
the Fast Fourier Transform performed with the open source library
\textit{fftw}~\citep{fftw}\footnote{\url{http://www.fftw.org}}.

We use the same value of $T$ as \citet{Pletsch1}. The Nyquist
frequency is $N/2T = 64\,\mbox{Hz}$ which determines a choice of
$N$. The relatively high value of the Nyquist frequency is required to
minimize the artificial binning noise keeping in mind that some
pulsars may be found at the second harmonic of the frequency.  We scan
over the parameter $\gamma$ from $0$ to $-1.6\times 10^{-11}$ with a
step equal to $-1\times 10^{-15}$. The range corresponds to the pulsar
characteristic age greater than $1$\,kyr. The step is selected taking
into account the computation time available for the project. The
value used introduces moderate loss of power for pulsations with
frequencies higher than $10$~Hz, see Eq.~3 of \citet{Ziegler:2008}.

The values of $f$ and $\dot{f}$ corresponding to the highest $P_m$ are
used as a starting point for the final coherent scan. The latter is
performed with maximization of the weighed $H$-test
statistic~\citep{Htest} defined as follows:

\begin{align}
\label{H}
H = \max_{1 \leq L
  \le20}\left[\sum_{l=1}^{L}\mid\alpha_{l}\mid^{2}-4(L-1)\right]\,,
\end{align}
where $\alpha_l$ is a Fourier amplitude of the $l$-th harmonic
\begin{align*}
\alpha_{l} &= \frac{1}{\varkappa}\sum_{a}w_{a}\exp^{-2\pi i l f \tilde
  t_a}\,,\\
\varkappa^2 &= \frac{1}{2}\sum_{a}w_{a}^2\,.
\end{align*}
 
The maximization is performed by a scan over $f$ and $\gamma$ with the
steps of $5\times 10^{-9}\,\mbox{Hz}$ and $2.5\times
10^{-18}\,\mbox{s}^{-1}$. The steps are chosen to ensure the coherence
over the whole observation time interval.  We take into account that a
combination of the pulsar's frequency and an inverse Fermi orbital
period $T_{LAT}$ may be found at the semi-coherent stage. Therefore,
the $H$-test is repeated three times starting from $f$, $f+1/T_{LAT}$
and $f-1/T_{LAT}$.

\begin{figure}[t]
\includegraphics[width=0.48\textwidth]{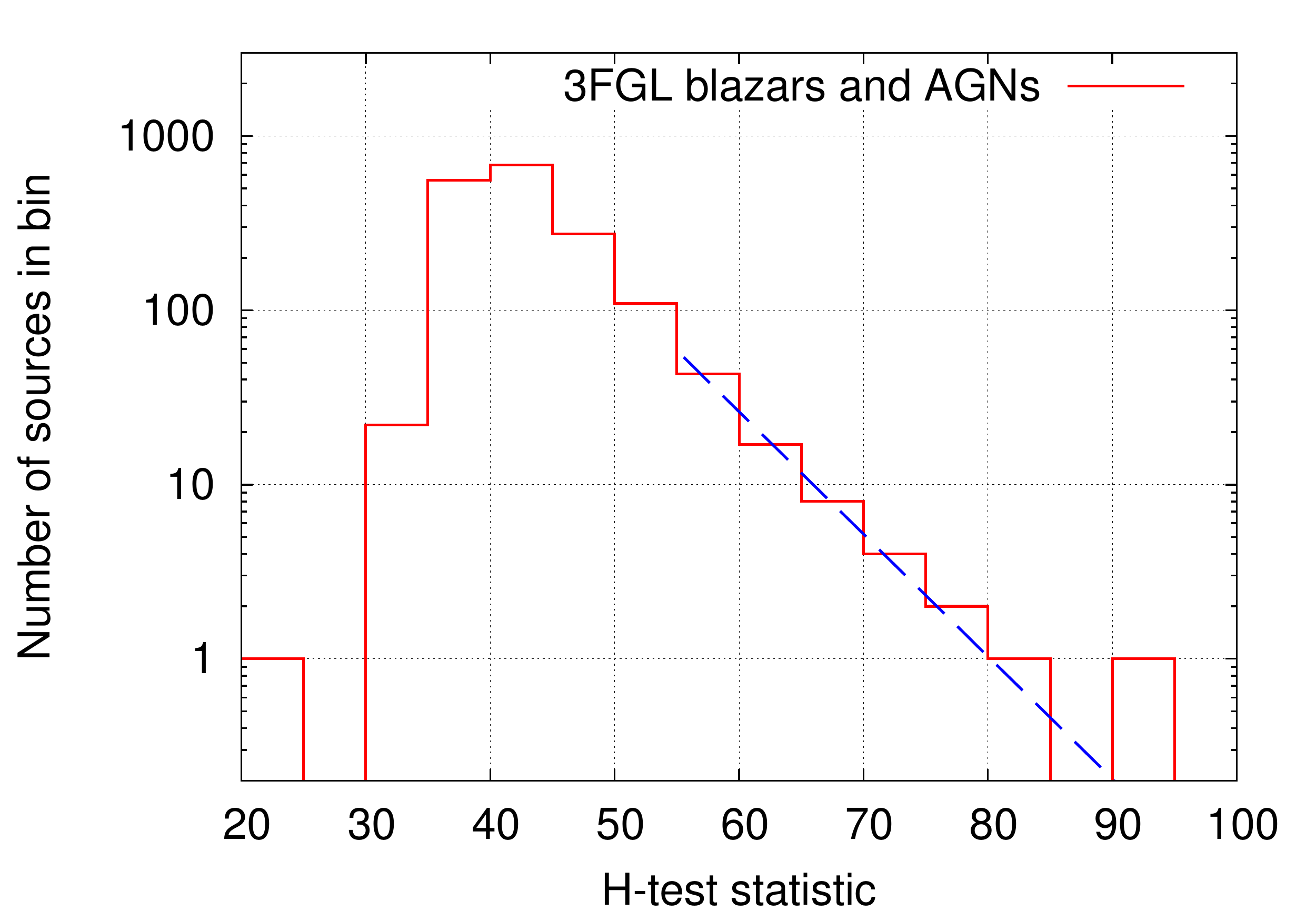}
\caption{The distribution of $H$-test statistic for 3FGL indentified 
blazars}
\label{histo}
\end{figure}

In this work an extensive scanning is performed and therefore a
theoretical distribution of $H$-test statistic for the null hypothesis
of non-pulsating object may not be used directly. We estimate the $H$
distribution using the results of the scan for the 1720 3FGL objects
identified as blazars, see Fig.~\ref{histo}. We extrapolate the tail
of the distribution with the exponential function and require that the
probability to have a single false candidate in the whole scan is less
than $5\%$. Thus we arrive to the threshold value $H_{th}=98$. Unlike
the rest of the work here we used identification information to select
blazars from all the sources. We note that this does not affect
detection uniformity since the same $H_{th}$ is then used for all
sources. Moreover, the procedure stays conservative in case of pulsar
contamination in the blazars sample. In this case the $H_{th}$ would
be overestimated and the probability of the false candidate appearance
even less than required.

\begin{deluxetable}{cccccc|cccccc}
\tabletypesize{\footnotesize}
\tablecaption{\label{catalog} A catalog of gamma-ray pulsars found in a
  blind search.}
\tablewidth{0pt}
\tablehead{
\colhead{no} & \colhead{3FGL name} & \colhead{$H$-} &
\colhead{$f$,} & \colhead{$\dot{f}, -10^{-13}$} & \colhead{age,} &
\colhead{Pulsar} & \colhead{l,} & \colhead{b,} &
\colhead{G,$10^{-11}$} & \colhead{Type} & \colhead{Ref.} \\ 
\colhead{~} & \colhead{~} & \colhead{test} & \colhead{Hz} & \colhead{Hz\,$s^{-1}$} &
\colhead{kyr} & \colhead{name} & \colhead{deg} & \colhead{deg} &
\colhead{erg\,cm$^{-2}$\,s$^{-1}$} & \colhead{~} & \colhead{~}
}

\startdata
1 & \object{J0007.0+7302$^\star$} & 658 & 3.16574810 & 36.1543 & 14 & \object{PSR J0007+7303} & 119.66 & 10.46 & 42.6 & Q & 1 \\ 
2 & \object{J0357.9+3206} & 485 & 2.25189503 & 0.6439 & 554 & \object{PSR J0357+3205} & 162.76 & -15.99 & 6.6 & Q & 1 \\ 
3 & \object{J0534.5+2201$^\star$} & 1750 & 29.72739783 & 3698.8692 & 1.3 &\object{Crab pulsar} & 184.55 & -5.78 & 147.2 & L & 2 \\ 
4 & \object{J0633.7+0632} & 866 & 3.36250736$^\dagger$ & 8.9973 & 59 & \object{PSR J0633+0632} & 205.10 & -0.93 & 12.4 & Q & 1  \\ 
5 &  \object{J0633.9+1746} & 11253 & 4.21755990$^\dagger$ & 1.9518 & 342 &  \object{Geminga pulsar} & 195.13 & 4.27 & 415.3 & Q & 3,4 \\ 
6 & \object{J0659.5+1414} & 199 & 2.59796057 & 3.7094 & 111 &  \object{Monogem pulsar} & 201.08 & 8.22 & 2.8 & L & 5 \\ 
7 & \object{J0734.7-1558$^\star$} & 189 & 6.44573114 & 5.1970 & 197 & \object{PSR J0734-1559} & 232.05 & 2.01 & 4.2 & Q & 6 \\ 
8 & \object{J0835.3-4510$^\star$} & 1353 & 11.18978060 & 156.0544 & 11 & \object{Vela pulsar} & 263.55 & -2.79 & 893.0 & L & 7 \\ 
9 & \object{J1028.4-5819} & 699 & 10.94042199$^\dagger$ & 19.2705 & 90 & \object{PSR J1028-5819} & 285.07 & -0.50 & 25.1 & L & 8 \\ 
10 & \object{J1044.5-5737} & 252 & 7.19264583 & 28.2328 & 40 & \object{PSR J1044-5737} & 286.57 & 1.15 & 13.6 & Q & 9 \\ 
11 & \object{J1048.2-5832} & 147 & 8.08368205$^\dagger$ & 62.7274 & 20 & \object{PSR J1048-5832} & 287.43 & 0.58 & 20.1 & L & 10 \\ 
12 & \object{J1057.9-5227} & 7561 & 5.07321957 & 1.5029 & 535 & \object{PSR J1057-5226} & 285.99& 6.64 & 29.0 & L & 10 \\ 
13 & \object{J1413.4-6205} & 371 & 9.11230510 & 22.9610 & 63 & \object{PSR J1413-6205} & 312.37 & -0.73 & 19.8 & Q & 9 \\ 
14 & \object{J1418.6-6058$^\star$} & 111 & 9.04346192$^\dagger$ & 138.3125 & 10 & \object{PSR J1418-6058} & 313.32 & 0.13 & 31.0 & Q & 1 \\ 
15 & \object{J1429.8-5910$^\star$} & 101 & 8.63231846$^\dagger$ & 22.7099 & 60 & \object{PSR J1429-5911} & 315.26 & 1.32 & 8.8 & Q & 9 \\ 
16 & \object{J1459.4-6053} & 313 & 9.69450009 & 23.7476 & 65 & \object{PSR J1459-6053} & 317.88 & -1.80 & 11.0 & Q & 1 \\ 
17 & \object{J1509.4-5850$^\star$} & 562 & 11.24546441 & 11.5913 & 154 & \object{PSR J1509-5850} & 319.98 & -0.62 & 10.3 & L & 11 \\ 
18 &\object{J1620.8-4928$^\star$} & 211 & 5.81616340 & 3.5480 & 260 & \object{PSR J1620-4927} & 333.90 & 0.39 & 18.0 & Q & 12 \\ 
19 & \object{J1709.7-4429} & 3543 & 9.75607885 & 88.5318 & 17 & \object{PSR J1709-4429} & 343.10& -2.69 & 131.5 & L & 13 \\ 
20 & \object{J1732.5-3130} & 1247 & 5.08792277 & 7.2603 & 111 & \object{PSR J1732-3131} & 356.31 & 1.02 & 14.9 & Q & 1 \\ 
21 & \object{J1741.9-2054} & 5503 & 2.41720733 & 0.9926 & 386 & \object{PSR J1741-2054} & 6.41 & 4.90 & 11.8 & L & 1 \\ 
22 & \object{J1809.8-2332} & 2790 & 6.81248050 & 15.9679 & 68 & \object{PSR J1809-2332} & 7.39 & -2.00 & 44.8 & Q & 1 \\ 
23 & \object{J1826.1-1256$^\star$} & 411 & 9.07223648$^\dagger$ & 100.0361 & 14 & \object{PSR J1826-1256} & 18.56 & -0.38 & 41.5 & Q & 1 \\ 
24 & \object{J1836.2+5925} & 756 & 5.77154961$^\dagger$ & 0.5005 & 1828 & \object{PSR J1836+5925} & 88.88 & 25.00 & 59.8 & Q & 1 \\ 
25 & \object{J1846.3+0919} & 282 & 4.43357094 & 1.9521 & 360 & \object{PSR J1846+0919} & 40.69 & 5.35 & 2.5 & Q  & 9 \\ 
26 & \object{J1907.9+0602} & 210 & 9.37783521 & 76.9151 & 19 & \object{PSR J1907+0602} & 40.19 & -0.90 & 31.9 & L & 1,14 \\ 
27 & \object{J1952.9+3253$^\star$} & 122 & 25.29478173$^\dagger$ & 37.4774 & 107 & \object{PSR J1952+3252} & 68.78& 2.83 & 15.1 & L & 15 \\ 
28 & \object{J1954.2+2836$^\star$} & 127 & 10.78634243$^\dagger$ & 24.6212 & 69 & \object{PSR J1954+2836} & 65.24 & 0.38 & 10.8 & Q & 9 \\ 
29 & \object{J1957.7+5034} & 349 & 2.66804343 & 0.5043 & 839 & \object{PSR J1957+5033} & 84.60 & 11.01 & 3.2 & Q  & 9 \\ 
30 & \object{J1958.6+2845} & 452 & 3.44356138 & 25.1308 & 22 & \object{PSR J1958+2846} & 65.88 & -0.35 & 9.9 & Q & 1 \\ 
31 & \object{J2021.1+3651} & 562 & 9.63902020$^\dagger$ & 89.0385 & 17 & \object{PSR J2021+3651} & 75.23 & 0.11 & 50.4 & L & 16 \\ 
32 & \object{J2021.5+4026$^\star$} & 122 & 3.76904980$^\dagger$ & 8.1591 & 73 & \object{PSR J2021+4026} & 78.23 & 2.08 & 88.3 & Q &  1 \\ 
33 & \object{J2028.3+3332} & 323 & 5.65907212 & 1.5557 & 576 & \object{PSR J2028+3332} & 73.37 & -3.00 & 6.4 & Q & 12 \\ 
34 & \object{J2030.0+3642} & 192 & 4.99678975 & 1.6231 & 488 & \object{PSR J2030+3641} & 76.13 & -1.43 & 4.6 & L & 17  \\ 
35 & \object{J2030.8+4416$^\star$} & 160 & 4.40392491$^\dagger$ & 1.2603 & 554 & \object{PSR J2030+4415} & 82.35 & 2.89 & 5.2 & Q & 12 \\ 
36 & \object{J2032.2+4126} & 106 & 6.98089488$^\dagger$ & 9.9452 & 111 & \object{PSR J2032+4127} & 80.22 & 1.02 & 16.0 & L & 1 \\ 
37 & \object{J2055.8+2539} & 807 & 3.12928985 & 0.4016 & 1236 & \object{PSR J2055+2539} & 70.69 & -12.53 & 5.5 & Q & 9 \\ 
38 & \object{J2140.0+4715$^\star$} & 121 & 3.53545109 & 0.2225 & 2520 & \object{PSR J2139+4716} & 92.64 & -4.04 & 2.3 & Q & 12 \\ 
39 & \object{J2229.0+6114$^\star$} & 152 & 19.36285105 & 294.8890 & 10 & \object{PSR J2229+6114} & 106.65 & 2.95 & 23.5 & L & 18 \\ 
40 & \object{J2238.4+5903} & 152 & 6.14486843$^\dagger$ & 36.5838 & 27 & \object{PSR J2238+5903} & 106.55 & 0.48 & 5.9 & Q & 1 \\ 
\enddata
\tablecomments{Frequency $f$ and spin-down rate $\dot f$ of gamma
  pulsations correspond to the epoch MJD 55225. Characteristic age is
  estimated as $-f/2\dot f$. The last six columns contain the object
  information from the literature: pulsar name, J2000 Galactic
  coordinates, Fermi LAT energy flux for $E>100$\,MeV\,\citep{3FGL},
  type (Q - radio-quiet, L - radio-loud) and a reference to the first
  identification of gamma pulsations. $\star$ new pulsars with respect to the
  catalog of the previous work~\citep{pscan1}. $\dagger$ pulsars 
detected at the second harmonic of the frequency.}
\tablerefs{\scriptsize
  (1) \citet{Abdo12}; (2) \citet{Crab}; (3) \citet{GemingaX}; (4)
  \citet{GemingaG}; (5)  \citet{Ma4}; (6) \citet{Saz}; (7)
  \citet{Vela}; (8) \citet{Abdo:J1028-5819}; (9) \citet{Parkinson69};
  (10) \citet{Thompson7}; (11) \citet{Fermi_2010ApJ}; (12)
  \citet{Pletsch1}; (13) \citet{Thompson11}; (14) \citet{Abdo:2010}; (15) \citet{J1952+3252}; (16) \citet{Halpern2}; (17) \citet{Camilo22}; (18) \citet{Halpern}.
}
\end{deluxetable}

\section{Results}
\label{sec:results}

There are 40 pulsars found in a blind search, see
Table~\ref{catalog} for the complete catalog. All the sources found in
this {\it Paper} were known previously as gamma-ray pulsars. These
include 25 radio-quiet and 15 radio-loud pulsars. There are 15 pulsars
detected at the second harmonic of their frequency -- these are marked
with a dagger in Table~\ref{catalog}. We note that all 25
pulsars from our previous blind search~\citep{pscan1} are also found
here. The blind search requires the threshold value of $H_{th}$ to be
relatively high in order to exclude possible false
detections. Meanwhile, 11 known gamma-ray pulsars, namely PSR
J0205+6449, PSR J0248+6021, PSR J1124-5916, PSR J1135-6055, PSR
J1422-6138, PSR J1746-3239, PSR J1803-2149, PSR J1813-1246, PSR
J1833-1034, PSR J1838-0537 and PSR J1932+1916, are detected with the
$H$-test statistic between $52.5$ and $98$. While the correct
frequency is found for these pulsars the identification is unfeasible
without prior knowledge and therefore they are not included in the
catalog.

\begin{deluxetable}{lcc}
\tablecaption{\label{tbl:KS} Radio-quiet and radio-loud pulsar population comparison}
\tablewidth{0pt}
\tablehead{ \colhead{Parameter} & \colhead{KS probability} & \colhead{AD probability}}
\startdata
P ($1/f$)& 0.8\% & 0.4\% \\
$\dot{P}$ & 52\% & 57\% \\
Age ($-f/2\dot f$) & 21\% & 13\% \\
Luminosity $(\sim f\dot f)$ & 3\% &2\%\\
Gamma energy flux & 12\% & 13\%\\
l & 99\% & 99\% \\
b & 99\% & 96\% \\
\enddata
\tablecomments{The KS-test and AD-test probabilities for comparison of
  radio-quiet and radio-loud pulsar distributions over period, its
  time derivative, age,
  spin-down luminosity, energy flux above 100 MeV and Galactic coordinates.}
\end{deluxetable}

\begin{figure}
\includegraphics[width=0.48\textwidth]{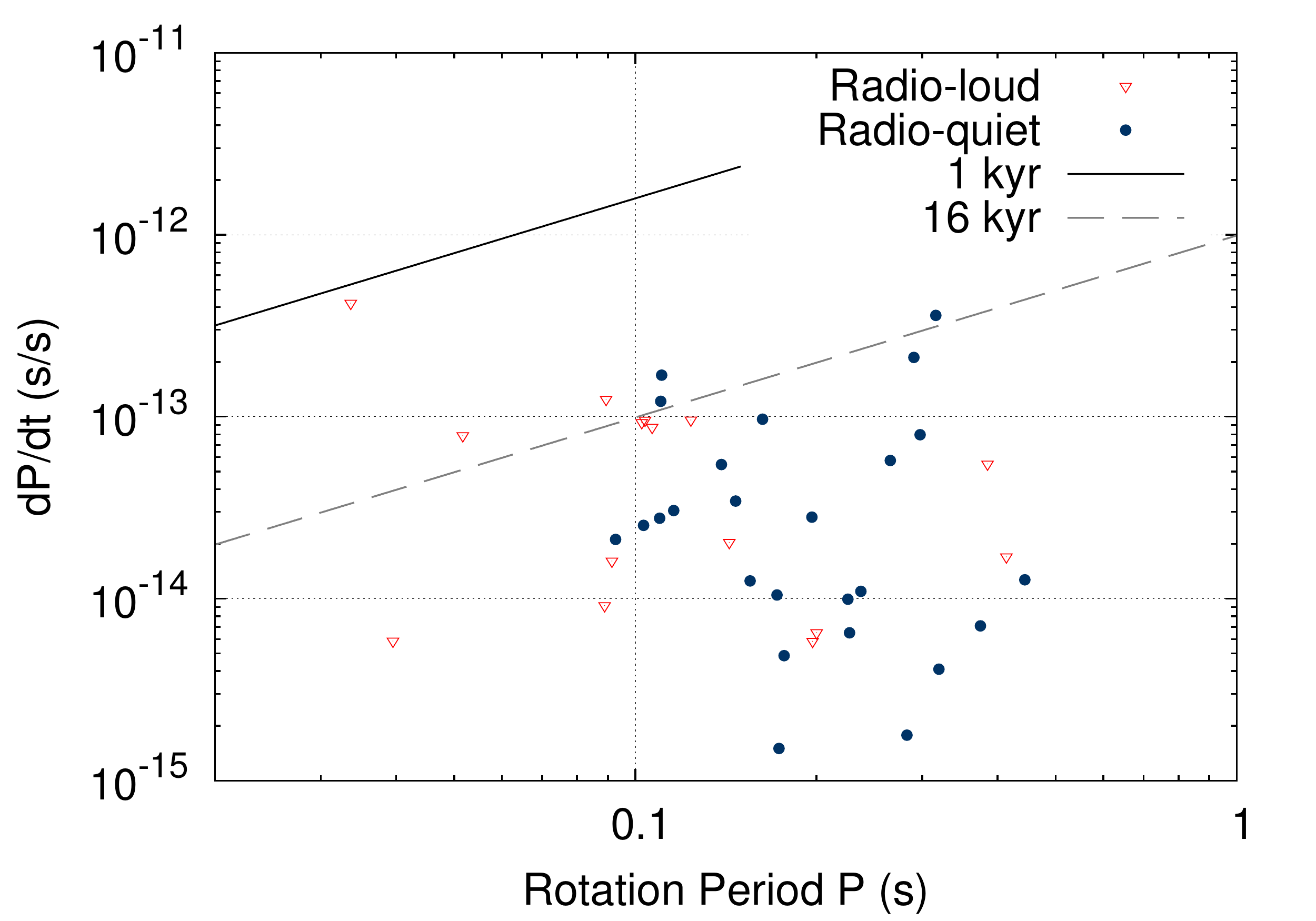}
\caption{ $P-\dot{P}$ plot for 40 pulsars found with a blind search in
  the present {\it Paper}. $P=\frac{1}{f}$ is a rotation period. The
  lines show maximum characteristics ages for present and previous
  blind searches.}
\label{pp}
\end{figure}

\begin{figure}
\includegraphics[width=0.48\textwidth]{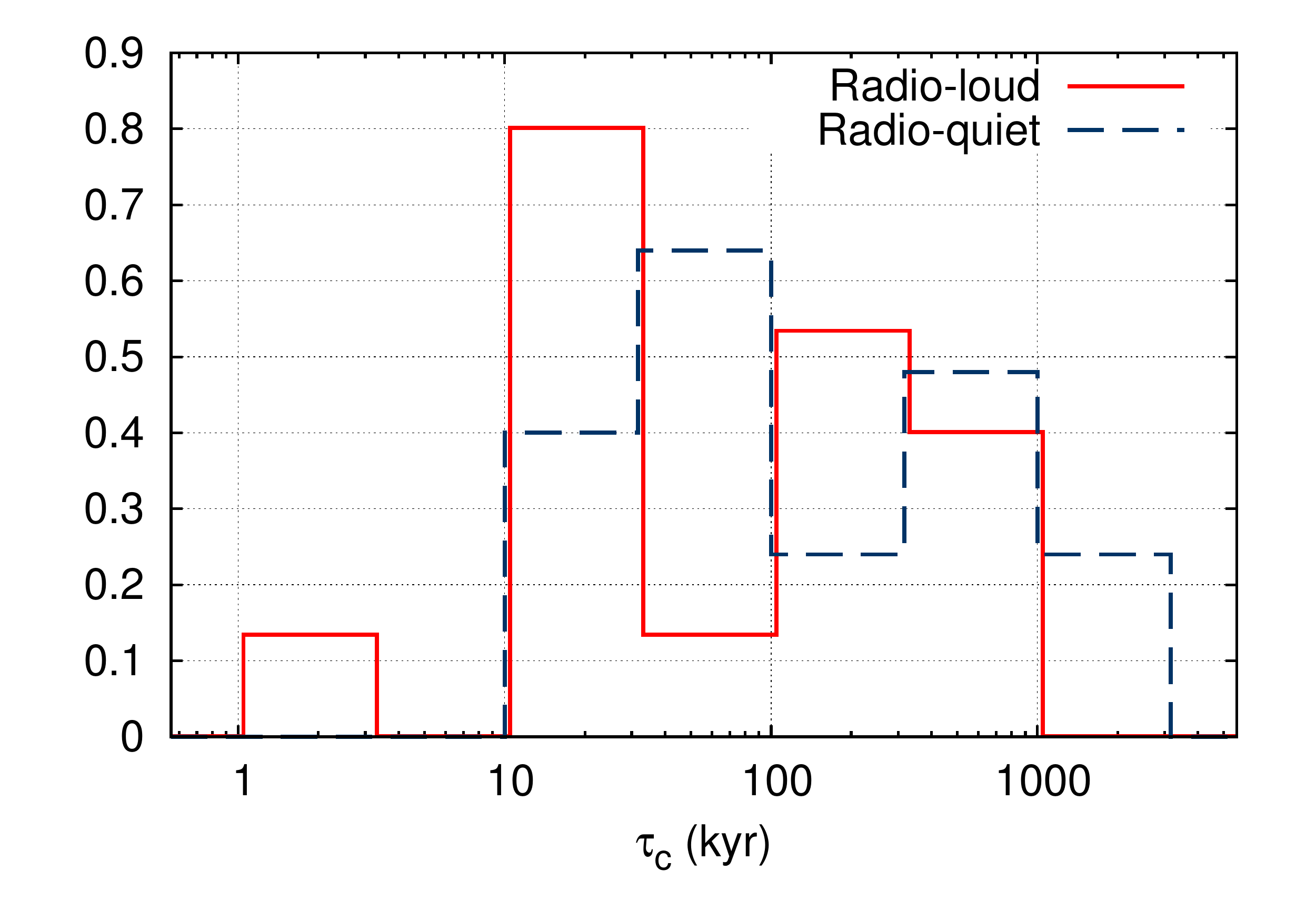}
\caption{Distributions of characteristic age $\tau_c=-\frac{f}{2\dot{f}}$
for radio-loud and radio-quiet pulsars. The two distrubutions are
compatible with KS probability $21\%$.}
\label{tau}
\end{figure}

\begin{figure}
\includegraphics[width=0.48\textwidth]{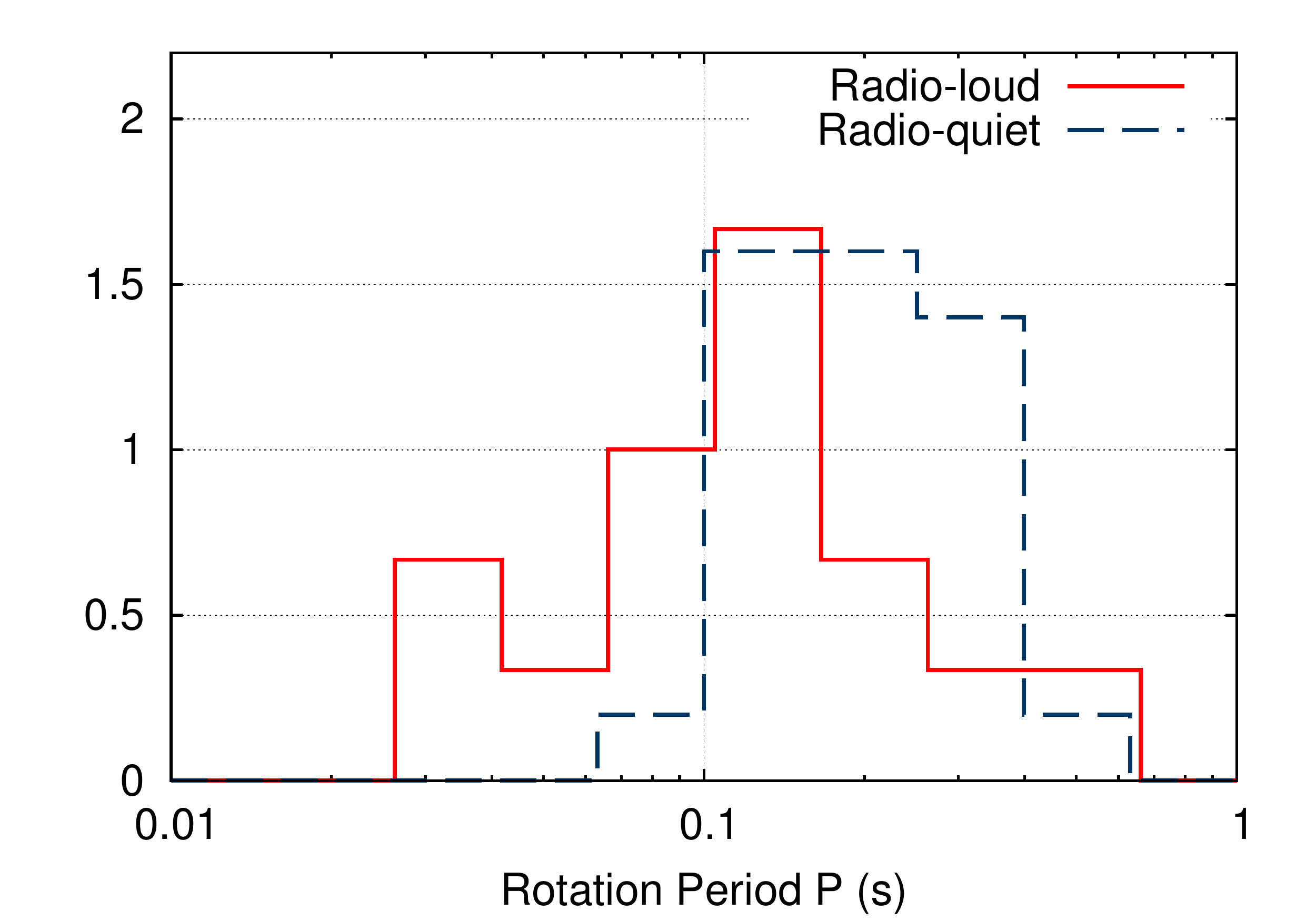}
\caption{Distributions of the rotation period $P$
for radio-loud and radio-quiet pulsars. The two distrubutions are
compatible with KS probability $0.8\%$.}
\label{p}
\end{figure}

\begin{figure}
\includegraphics[width=0.48\textwidth]{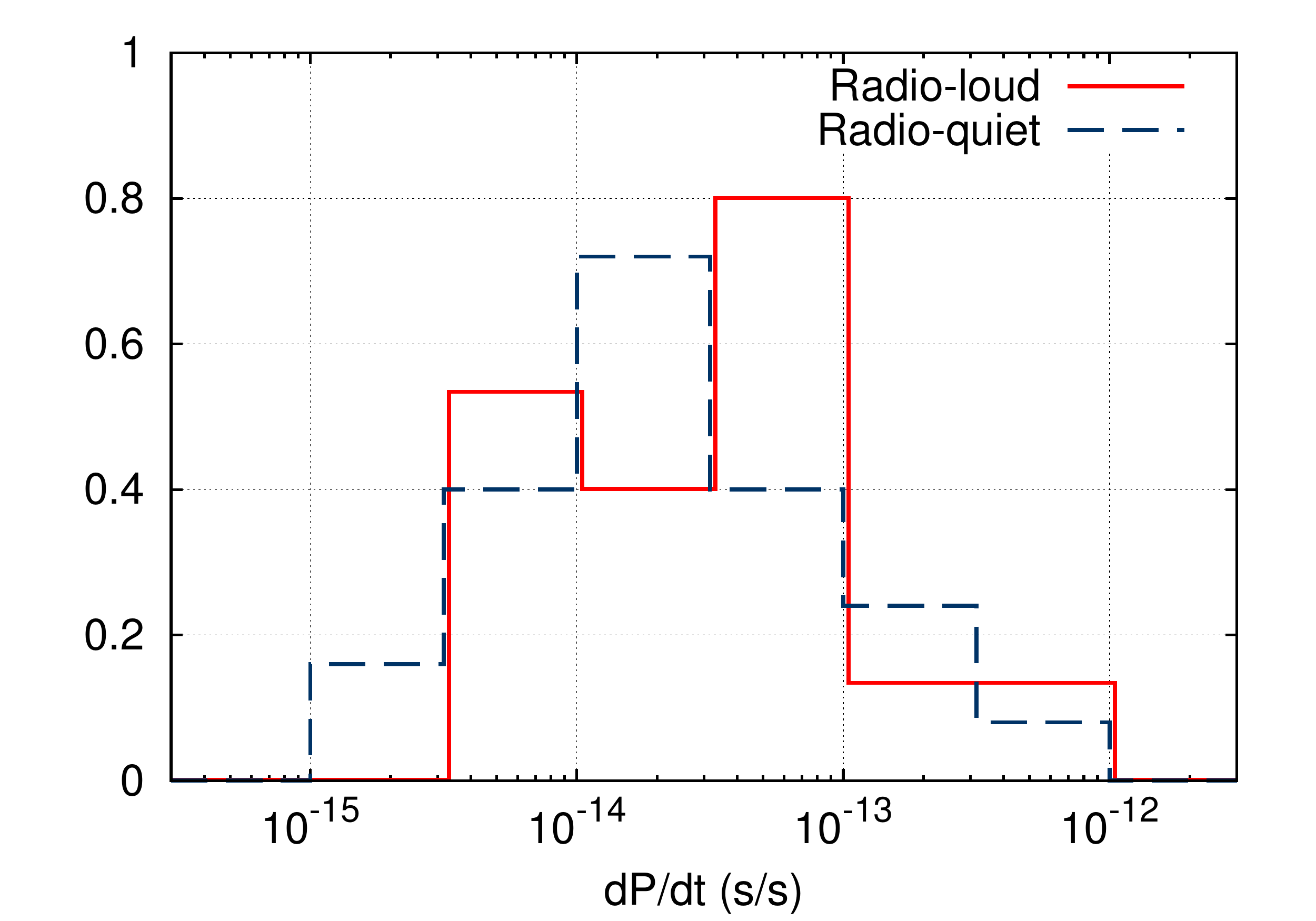}
\caption{Distributions of $\dot{P}$
for radio-loud and radio-quiet pulsars. The two distrubutions are
compatible with KS probability $52\%$.}
\label{pdot}
\end{figure}

\begin{figure}
\includegraphics[width=0.48\textwidth]{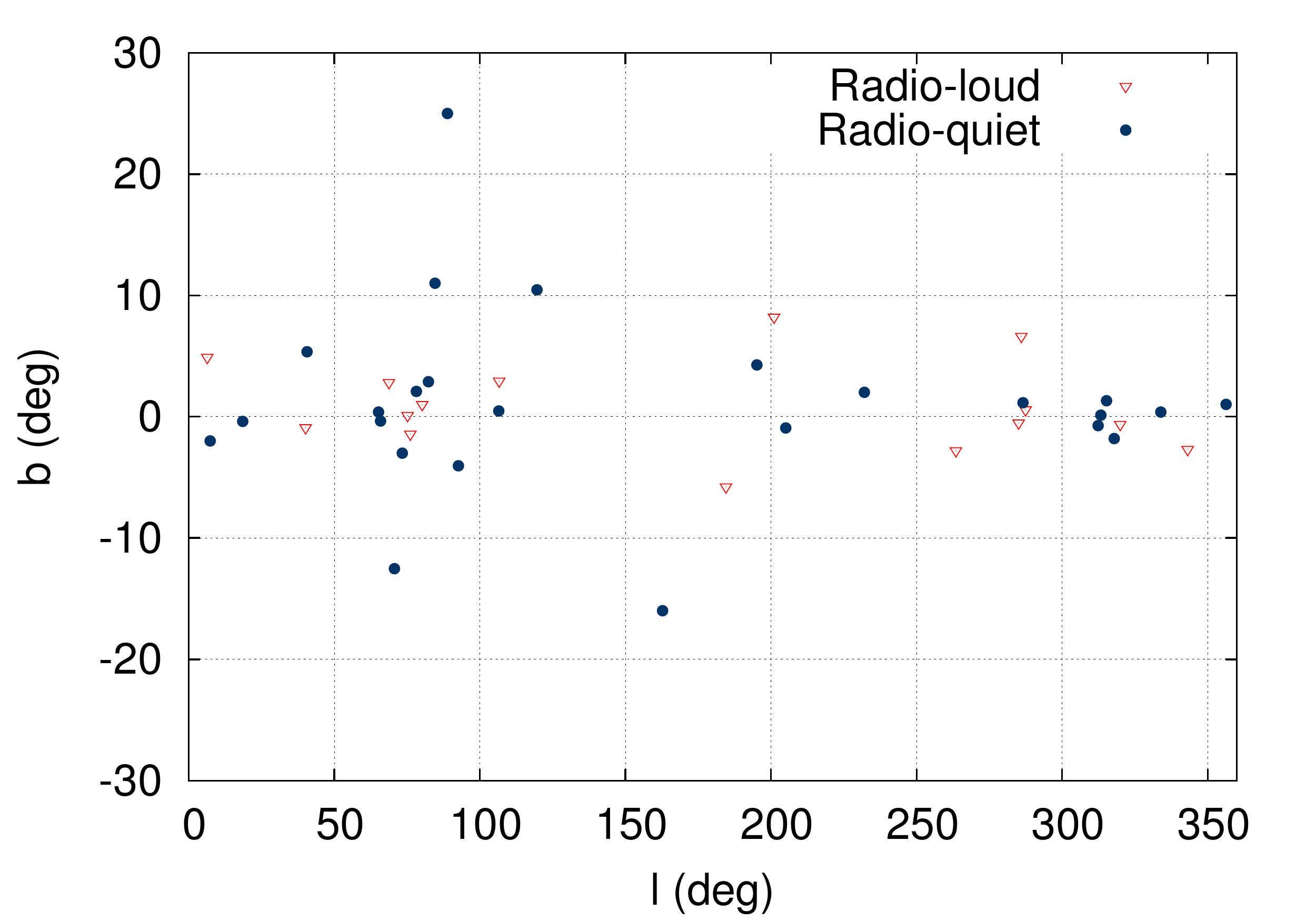}
\caption{Galactic coordinates of radio-loud and radio-quiet gamma-ray pulsars}
\label{lb}
\end{figure}

\begin{figure}
\includegraphics[width=0.48\textwidth]{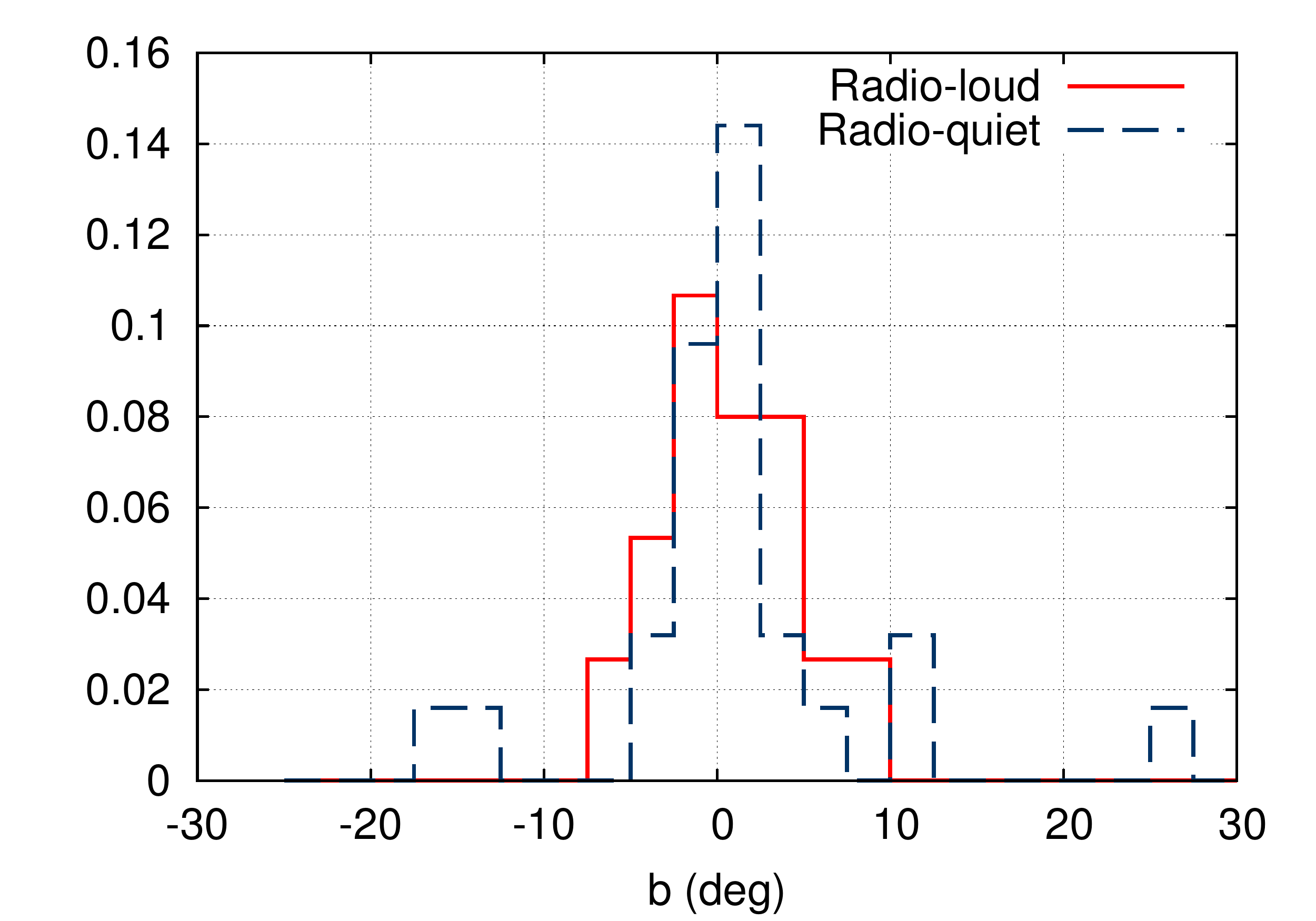}
\caption{Distributions of Galactic latitude $b$ 
for radio-loud and radio-quiet pulsars. The two distrubutions are
compatible with KS probability $99\%$}
\label{b}
\end{figure}

We compare the distributions of the observed parameters for the
radio-loud and radio-quiet pulsars with the Kolmogorov-Smirnov (KS)
and Anderson-Darling (AD) tests using {\it R}~\citep{R} and {\it kSamples}
package~\citep{kSamples}. We summarize the results in
Table~\ref{tbl:KS} and Figures~\ref{pp}-\ref{b}. The tests were
performed . There are no statistically significant differences in
characteristic age, $\dot{P}$, spin-down luminosity, gamma-ray
luminosity and Galactic coordinates. The rotation period histograms
are marginally different with pre-trial AD-probability of $4\times
10^{-3}$. The post-trial significance is only $5\%$ taking into
account that 6 independent tests are performed with 2 different
statistical methods (AD and KS).

Based on the general agreement of the observed parameters of
radio-loud and radio-quiet pulsars, one may assume that the chance
probability of the particular pulsar to enter the blind search catalog
does not depend on its radio emission properties. Therefore the
fraction of the radio-quiet pulsars in the whole population of the
gamma-ray pulsars may be estimated with the corresponding fraction in
the catalog:
\begin{equation}
\label{epsilonRQ}
\epsilon_{RQ}=\frac{N_{RQ}}{N_{RQ}+N_{RL}} = 0.63\pm 0.08 ~\mbox{(68\%\,CL)},
\end{equation}
where $N_{RQ}$ and $N_{RL}$ are numbers of radio-loud and radio-quiet
non-recycled pulsars correspondingly. The result is in a good
agreement with our previous work~\citep{pscan1}.

Given that Fermi-LAT has observed 61 radio-loud pulsars and considering
$\epsilon_{RQ}$ in its general sense, we predict that there are about
104 radio-quiet pulsars within the Fermi-LAT sources for which the
pulsations are detectable if the precise position and ephemerides are
hypothetically known. Therefore, within these sources there are more
than 50 radio-quiet pulsars with still undiscovered pulsed
emission. These pulsars may be tracked when more gamma-ray data are
available or with the future breakthroughs in blind search techniques.

Finally, let us compare our results with the published predictions of
the emission models. The radio-quiet and radio-loud pulsar
distribution with age are statistically compatible, counter to the
prediction of the PC models that the radio-quiet fraction depends on
age. More specifically, the $\epsilon_{RQ}^{PC} \le 0.53$ estimated
for PC models and even smaller value at ages higher than 100
kyr~\citep{Gonthier:2002} are in tension with our results. On the other
hand, the $\epsilon_{RQ}^{OM}=0.65$ estimated in the OM
models~\citep{Perera:2013wza} is in perfect agreement with
Eq.~\ref{epsilonRQ}. While the catalog covers nearly half of
the known radio-quiet pulsars, there is no indication of the evolution
of the radio-beam solid angle proposed in \citet{Ravi:2010sm}.

{\bf Acknowledgments}

We are indebted to A.G.~Panin for numerous inspiring discussions. We
thank V.S.~Beskin, M.S.~Pshirkov and S.V.~Troitsky for useful comments
and suggestions. We are obliged to the anonymous referee of The
Astrophysical Journal Letters for suggesting more efficient analysis
technique as a comment to our previous paper and to the anonymous
referee of The Astrophysical Journal work for suggesting many
improvements. The work is supported by the Russian Science Foundation
grant 14-12-01340. The analysis is based on data and software provided
by the Fermi Science Support Center (FSSC). We used SIMBAD
astronomical database, operated at CDS, Strasbourg, France. The
numerical part of the work is performed at the cluster of the
Theoretical Division of INR RAS.

\end{document}